\documentclass{emulateapj}

\begin{document}

\title{Determining atmospheric conditions at the terminator of the hot-Jupiter HD209458\MakeLowercase{b}}

\author{David K. Sing\altaffilmark{1}}
\author{A. Vidal-Madjar\altaffilmark{1}} 
\author{A. Lecavelier des Etangs\altaffilmark{1}}
\author{J.-M. D\'{e}sert\altaffilmark{1}}
\author{G. Ballester\altaffilmark{2}}
\author{D. Ehrenreich\altaffilmark{3}}

\altaffiltext{1}{Institut d'Astrophysique de Paris, CNRS; Universit\'{e} 
Pierre et Marie Curie, 98 bis bv Arago, F- 75014 Paris, France; sing@iap.fr}

\altaffiltext{2}{Lunar and Planetary Laboratory, University of Arizona, 
Sonett Space Science Building, Tucson, AZ 85721-0063, USA}

\altaffiltext{3}{Laboratoire d'astrophysique de Grenoble, Universit\'{e} 
Joseph-Fourier, CNRS (UMR 5571), BP 53 F-38041 Grenoble cedex, France}

\submitted{Accepted to Ap.J May 6, 2008}

\begin{abstract}

 We present a theoretical model fit to the {\it HST}/STIS optical transit transmission spectrum
of HD209458b.  In our fit, we use the sodium absorption line profile along with
the Rayleigh scattering by H$_{2}$ to help determine the average temperature-pressure profile at
the planetary terminator, and infer the abundances of atomic and molecular species.
The observed sodium line profile spans an altitude range of $\sim$3,500 km, corresponding to
pressures between $\sim$0.001 and 50 mbar in our atmospheric models.
We find that the sodium line profile requires either condensation into sodium sulfide or ionization, 
necessary to deplete atomic sodium only at high altitudes below pressures of $\sim$3 mbar.
The depletion of sodium is supported by an observed sudden abundance change, from 2 times solar 
abundance in the lower atmosphere to 0.2 solar or lower in the upper atmosphere.
Our findings also indicate the
presence of a hot atmosphere near stratospheric altitudes corresponding to pressures of 33 mbar, consistent
with that of the observed dayside temperature inversion. In addition, we find a separate 
higher altitude temperature rise is necessary within both the condensation and ionization models, 
corresponding to pressures below $\sim$0.01 mbar.  
This hot higher altitude temperature indicates that absorption by atomic sodium can potentially 
probe the bottom of the thermosphere, and is possibly sensitive to
the temperature rise linked to atmospheric escape.

\end{abstract}


\section{Introduction}

Early theoretical transit models of hot-Jupiter exoplanetary
atmospheres \citep{SS00,Brown01,Hubbard01} contained strong sodium absorption
signatures, predicted to be as high as 0.2\% for clear cloudless
atmospheres.  The original measured Na absorption signature in
HD209458b from medium resolution {\it HST} data \cite{Charby} was,
however, $\sim$3$\times$ smaller than initial predictions.
Additionally, \cite{Bar07} has also identified 
a potential broad band Na signature from photometry of low resolution STIS data in the same planet,
taken several years later \citep{Knut}.  
These observations have lead to multiple
theories to explain the small Na signature.  The most widely cited
explanations of reduced Na absorption include: high clouds or hazes,
ionization, low global Na abundance, rainout, as well as non-local
thermal equilibrium effects \citep{Charby,Bar02,Bar07,Fortney03}.  
High opaque clouds could potentially mask absorption
features by effectively cutting transit information for altitudes lower
than the cloud deck \citep{SS00,Fortney03}.  The small atmospheric signature has
made detection from ground-based telescopes difficult, with no Na
detection currently reported for HD209458b after multiple
attempts \citep{Narita,Arribas}.

In the grazing line of sight of transit geometry, the measured effective altitude 
depends upon the wavelength dependant opacities of the atmospheric
constituents giving a transit transmission spectrum its altitude sensitivity
(see relevant equations in \citealt{Lecavelier08a}).
The effective altitude is also dependant upon the abundance of
absorbing atmospheric particles, as well as the atmospheric pressure.
As the abundances are not in general a priori known, this leads to atmospheric models which have degenerate
solutions between modeled abundance and pressure, where an atmospheric transit model with increased
abundance is identical to models with lower abundances and higher pressures.

Here we interpret the optical {\it HST} STIS transmission spectra of HD209458b of
\cite{Sing08a}.  The spectra consists of STIS data obtained during planetary 
transit at low and medium spectral resolutions.
The two datasets are combined to extend 
the measurements over the entire optical regime, providing a
way to simultaneously measure, in the case of atomic Na, 
both the narrow line core and wide line wing 
absorption as well as other absorbing or scattering species.  Since the 
Na line core and wings are optically thick at different
atmospheric heights, a detailed analysis of the Na line profile in conjunction with
other observed species can potentially break the model degeneracies and reveal
atmospheric structure such as the altitude-temperature-pressure (z-T-P) relation,
abundance, and the possible presence and height of planetary clouds.
In this paper, we focus mainly on the resulting temperature pressure profile, as well
as the Na line itself.  We briefly describe the observations and resulting spectra in \S2, present our
transit model fits in \S3, report on an observed temperature inversion in \S3.3, and discuss our results
in \S4 \& 5. In addition to this work, \cite{Lecavelier08b} provides a detailed 
description of H$_{2}$ Rayleigh scattering while \cite{desert08} goes into modeling details of TiO, 
VO, and other molecular species.

\section{Observations}

The data reduction, procedures, and various tests resulting in the transmission spectra
used in this analysis are covered in detail by \cite{Sing08a}, though we 
briefly describe them here for completeness.
The {\it HST}/STIS optical transit data for HD209458b consist of data at low
resolution covering $\sim$3,000-10,000{\AA} with the G430L and G750L gratings
and medium resolution data from the G750M grating covering 5,813-6,382{\AA}.

With the medium resolution data, \cite{Sing08a} measured the Na absorption using
the double differential photometric method of \cite{Charby}, as well as a
double differential spectroscopic evaluation.  For the differential
photometric measurement, \cite{Sing08a} selected 23 wavelength bands from 4 to 100
\AA\ centered on the Na~D lines, \cite{Charby} used three different
wavelength bands, and compared the absorption to that of regions
red-ward and blue-ward of the Na~D line region.  The resulting
photometric Na absorption profile (see Sing et al. 2008, Fig. 3) matches
the original results of \cite{Charby} and, in addition, reveals the strong
absorption from the Na~D line centers. 

To analyze the low and medium resolution {\it HST} spectra together, \cite{Sing08a} used a
limb-darkened corrected transit spectral ratio to probe the full
wavelength dependence of atmospheric absorption features (see Fig. 1 \& Fig. 2).  From this
analysis, the Na~D wavelength region shows an absorption plateau between
$\sim$5600-6700 {\AA}, which is $\sim$0.045\% above the minimum absorption
level observed (see Fig. 1).  Note that the wavelength range of the medium
resolution data, originally used to detect atmospheric Na \citep{Charby}, 
is entirely contained within this broad plateau
absorption region.  The total absorption seen in
the Na~D line cores, therefore, is the result from the medium
resolution data, which is the absorption above the plateau region
measured to be $\sim$0.065\%, {\it plus} the absorption level of the plateau
region itself.  This total Na core absorption level is $\sim$0.11\%, within a 4.4{\AA} band,
and is much closer to the original predictions \citep{SS00,Brown01},
indicating that a large extent of the atmosphere is being probed
during transit.  
Additionally, the Na~D doublet is resolved
in our resulting spectra which shows two strong narrow
absorption lines (see Fig. 2).

\subsection{Observed Na line profile}

The observed Na~D transit line profile (Fig. 1), with it's narrow
absorption core and broader plateau-like shape, is in stark contrast
to cloudless theoretical models of
HD209458b \citep{SS00,Brown01,Fortney03,Bar07} which have smoothly decaying
Lorentzian ``Eiffel Tower'' shaped line wings.  This more complicated
observed line profile suggests a very low quantity of Na is present in
the middle atmosphere and much larger quantity in the lower
atmosphere, pointing to either high altitude Na condensation into Na$_2$S or
alternatively, depletion via ionization.

Previous studies have predicted models of HD209458b which
included Na$_2$S clouds \citep{Iro05}, as high altitude temperatures on
the night side reach levels cool enough for condensation of atomic
Na \citep{Lodd99,Show02}.  Sodium sulfide would effectively
decrease the amount of atomic sodium at altitudes above the
condensation temperature, creating a stratified-like atmospheric profile
depleted of atomic Na only at those high altitudes.  The transit 
signature with such a layered Na profile
produces a broad plateau shaped line wing along with narrow absorption
cores rising above the plateau to higher levels \citep{Iro05}.  
Na$_2$S would have to be transparent, since they are not observed in absorption during
transit at the limb.  If Na$_2$S or other clouds/hazes were opaque in absorption during transit
at the condensation altitude, significant absorption would extend over a very broad 
wavelength range \citep{Fortney05}, masking the low resolution absorption features we observe.
Such a flat, largely featureless transmission spectra has recently been observed
in HD189733b, the result of high altitude haze \citep{Pont08, Lecavelier08a}.

An alternative method to deplete neutral atomic sodium is by ionization \citep{Fortney03}.
In this scenario, UV photons between wavelengths of 1,107 and 2,412 {\AA} (corresponding to 
11.2 and 5.14 eV respectively) photoionize sodium along the transit line of sight until recombination
matches ionization.  \cite{Fortney03} found a sharp boundary at the limb between higher atmospheric regions of complete
ionization and lower regions with no ionization.  Such a model could explain the observed Na line shape, with a narrow Na absorption 
core rising above broad line wings, if the atmosphere suddenly becomes ionized above a certain altitude.

\section{Model Fits to the transit transmission spectra of HD209458b}

We modeled the atmospheric transmission spectra calculating the
transit signature based on a theoretical model \citep{dav} containing absorption by
Na, H$_2$ Rayleigh scattering, TiO and VO.  The model
output absorption spectrum can then be directly compared to our
limb-darkened corrected transit absorption spectrum. 
We calculated the optical depth along a grazing line of sight through the planet's atmosphere, 
and calculated the absorption depth at a given wavelength using 
equations given in \cite{Lecavelier08a}. For the Na D 
lines, we used collision-broadened line-shapes calculated using 
Equations of \cite{Iro05}. Concerning atomic sodium, 
ultimately the free parameters to fit in the transit calculation 
include the atmospheric temperature profile as a function of altitude, and the mixing ratios of 
Na below and above the condensation or ionization level (see Fig. 3). 

We found two families of solutions, with each giving similarly good fits to the data.  In one model, Na is
allowed to condense into sodium sulfide at the chemical equilibrium temperature, while in the other model neutral Na is depleted at high altitudes,
presumably by ionization, below an arbitrary pressure.

\subsection{Condensation model}
\subsubsection{Temperature-Pressure profile}

For the calculation,
we fit various simplified T-P profiles in which
the temperature is taken as a linear function of the altitude and
pressure is calculated using hydrostatic equilibrium.  
Note that it is not specifically the T-P profile shape itself which is important in our fitting of
the data, but rather the different temperature and pressure regimes imposed by the observations.  At
our signal-to-noise levels, the general variation of the T-P profile in between the different atmospheric regimes
we find (hot at high pressure, cool at middle pressures, and hot at low pressures) is only loosely constrained.
A more complicated T-P profile would add undesired additional parameters into the fit, leading us to choose
as simple a T-P profile as possible.

We found it necessary to have a minimum of three variable points in the
T-P profile. The temperature as a function of the altitude, $T(z)$, is
thus composed of two linear functions below and above the altitude
$z_{m}$, where $z_{m}$ is the altitude of the minimum temperature,
$T_{m}$.  We took the lower reference altitude, $z_{s}$, for the
altitude corresponding to the mean planetary radius \citep{Knut}, and the
altitude $z_{th}$ is the fixed position, at high altitude, needed to
define the second part of the $T(z)$ linear function. We used $z_{th}$
corresponding to the altitude of the absorption depth of 1.55\%, which
corresponds to the mean absorption depth in the core of the sodium
doublet.  The temperature and pressure at the altitude $z_{s}$, the
altitude $z_{m}$ of minimum temperature $T_{m}$, and $T_{th}$ are
the five free parameters defining the T-P profile.  The sodium
abundance in the lower and the upper atmosphere are the two other free
parameters in the fit. 

In our condensation fit we found:
(1) a high temperature of 2,200$\pm$260 K at 33$\pm$5 mbar  ($\sim$500 km)
mainly constrained by Rayleigh scattering \citep{Lecavelier08b} seen at 
wavelengths shorter than $\sim$5,000 {\AA}, 
(2) a low temperature of 420$\pm$190 K at 0.63$\pm$0.27 mbar needed to explain the plateau at 0.045\% ($\sim$1,500 km)
above the minimum absorption level as a Na abundance change due to condensation, and finally 
(3) a high altitude ($\sim$3,500 km) high temperature of 1,770$^{+500}_{-570}$ K at 0.0063$\pm$0.003 mbar needed to 
fit the strong peak and narrow width of the cores of the Na~I doublet.

The pressure-temperature gradient we find when imposing condensation is close to and consistent (within 1$\sigma$)
of an adiabatic gradient.  A temperature gradient which is adiabatic or less is needed to be stable against 
convection, and solutions imposing adiabatic variations give acceptable fits to the data (an adiabatic fit is shown in Fig. 3).
We ultimately, however, choose to quote values found from the more general solution, as the linear temperature-altitude relation 
we assume includes adiabatic gradients and the general fit is located at a well behaved minimum $\chi^2$.  Although the
best fit solution is slightly superadiabatic, the relatively low quality of the data still allows for a wide range of T-P gradients and
the adiabatic-like profile found from our general fit indicates that an adiabatic behavior is likely at those altitudes.

\subsubsection{Atmospheric Na Condensation, TiO \& VO}

Our model contained two different amounts of atomic Na having separate variable
parameters for the mixing ratios of the middle and lower
atmospheres.  The pressure boundaries (or condensation pressure) where
the quantity changes occur is taken to be at an altitude where the
temperature profile crosses the Na$_2$S condensation curve.  In our
fits, we found a Na mixing ratio of 3.5$^{+2.9}_ {-1.9}\times10^{-7}$
in the middle atmosphere and 4.3$^{+2.1}_{-1.9}\times10^{-6}$ in the
lower atmosphere, corresponding to 0.2 and 2 times solar abundance.

\cite{Fortney05} estimated that
the optical depth of Na$_2$S condensates in transit geometry could be optically thick for HD209458b ($\tau\sim$0.73) for an atmosphere
of solar composition and cloud base at 30 mbar.  Given the fit T-P profile, we find a Na$_2$S cloud base of $\sim$3 mbar.
Lower pressures or larger particle sizes would provide a natural explanation for transparent clouds, as the condensate optical
depth is proportional to the pressure and inversely proportional to the particle size \citep{Marley2000}.  Given our determined Na abundances, the mixing ratio
of Na$_2$S we find is (2$\pm1$)$\times10^{-6}$, which matches the assumed value of \cite{Fortney05}.
Scaling the estimated optical depth of \cite{Fortney05}, a cloud base at 3 mbar gives $\tau\sim$0.073, indicating that transparent Na$_2$S 
condensates are plausible.  

The continuum blue-ward of the Na~D lines, between $\sim$4,000 and 5,500
\AA, shows a slightly lower absorption than the red-ward region,
7,000-8,000 \AA, and a significant absorption `bump' around $\sim$6,250\AA\ which 
appears in both resolutions.  
Our fits indicate that the red-ward absorption levels (7,000-8,000 {\AA})
are likely due to TiO and VO at stratospheric heights, being largely confined to altitudes
below our Na condensation level (also see \citealt{desert08}).  The TiO and VO features mask part of the long wavelength
Na line wing, though not the short wavelength side, providing strong constraints on
the abundance and altitude of TiO and VO.  The hot low altitude temperatures we find from Rayleigh scattering
are consistent with the presence of TiO and VO, as both species begin to condense out
of the atmosphere at temperatures below $\sim$1,800 K. 

The low-altitude pressure assumed here is dependent on the detection of
Rayleigh scattering \citep{Lecavelier08b}, though 
multiple interpretations of the absorption rise at those wavelengths
exist \citep{Lecavelier08b,Ballester07,Bar07}.  Ultimately, it may be difficult
without Rayleigh scattering to assign an absolute pressure scale to
these measurements which are inherently differential. 
The effective planetary radius seen during transit is sensitive to the quantity of
the atmospheric pressure times the abundance of absorbing species (see 
equations of \citealt{Lecavelier08a}), making 
an atmospheric fit at higher altitude possible by simply increasing the absorbers abundance.
The advantage of detecting Rayleigh scattering by H$_2$ when interpreting these models
is that the abundance of H$_2$ is known, constituting the bulk of the atmosphere, which thereby fixes
the pressure, and defines the pressure scale for the rest of the transmission spectrum. 
Our adoption of Rayleigh scattering here is based on the goodness of
fit to the data and the necessity to probe a large pressure range, 
given the total Na profile observed.

Since the low-altitude pressure is an upper limit
constrained by Rayleigh scattering, a lower pressure
would have an effect on the lower temperature profile, increase the
lower Na mixing ratio, and decrease the condensate pressure determined
in our transit fit.  Whatever the true pressure $P_{s}$ or additional
absorption sources, however, the basic atmospheric characteristics of Na
determined here would remain the same.  Namely, the middle
atmosphere has a greatly reduced amount of Na with a temperature below
Na$_2$S condensation, while the lower and higher atmospheres are
warmer than the condensation temperature with the lower atmosphere
containing a solar-like Na abundance and higher one a
significantly reduced Na abundance.

\subsection{Photoionization model}

Given the large uncertainties modeling hot-Jupiter Na photoionization, we choose not to employ a complex 
ionization model, but rather follow the results of \cite{Fortney03} and assume in our model that 
neutral Na can be suddenly ionized above a given altitude leading to the observed Na depletion.  
The ionization depth found by \cite{Fortney03} is around 1.7 mbar, close to the pressure we find for a Na
abundance change in the condensation model, indicating the effects could be important at those pressures
and such a model is viable.

A model with Na ionization effectively lifts the constraint of the temperature profile dropping below the Na
condensation curve, which the condensation model uses to deplete atomic Na.  Effectively, a model with Na 
ionization leaves a wide range of temperatures possible
in the middle atmosphere (corresponding in our model to $T_{m}$) as the collisional broadened Na profile at those
altitudes (1,000 - 2,000 km) contains no temperature information.  The corresponding H$_2$ Rayleigh scattering slope 
(below $\sim$3,400 {\AA}) is at the edge of the spectrum and also provides no useful middle atmospheric temperature constraints.  Within the framework of 
equilibrium chemistry, the temperature $T_{m}$ is likely below
that of the TiO/VO condensation curve and above that of the Na$_2$S condensation curve.
However, Rayleigh scattering still implies a high temperature of 2,200$\pm$260 K at 33$\pm$5 mbar and
the deep narrow Na cores suggest high altitude temperatures of around 1,770$^{+500}_{-570}$ K.
Without a strong constraint on the middle atmosphere temperature, T-P profile fits are dominated
by these upper and lower atmospheric constrains, leading to anomalously high temperature T-P profiles fit to be everywhere above 2,000 K.  
Though technically good fits to the data, such profiles can be ruled out as the temperatures would
be everywhere higher than the dayside temperature inversion seen with Spitzer \citep{Knut08,Burrows07}.
Additionally, the lack of observed Ca I at $\lambda$4,227 {\AA} in the low resolution data further suggests 
temperatures reach values lower than that required for Ca-Ti and Ca-Al-bearing condensates \citep{Lodd06}.

Though the current data lacks enough constraints for a unique T-P profile solution in the photoionization model, 
the ionization pressure, $P_{ion}$, can be estimated given the likely range of temperatures.  We fit various isothermal models fitting $P_{ion}$ 
as free parameter along with the Na abundances (taken to be constant) above and below that pressure.  The 
resulting ionization pressure varies from 0.5 mbar to 5 mbar for temperatures of 1,100 and 2,000 K respectively.
Higher temperature fits result in smaller upper-atmospheric Na abundances and larger Na abundance contrasts.
The abundances found in the lower temperature fits are in general agreement with those found in the Na condensation model, while
the upper temperature fits find much smaller upper-abundances, with the 2,000 K model giving an upper Na abundance 180 times less than solar. 
Thus, a Na abundance drop by a factor of ten can be considered as a lower limit, and may be much larger depending on the depletion
mechanism.

\subsection{High altitude temperature inversion}

A $\sim$0.11\% absorption of Na indicates that our transmission spectra probes an altitude range of
$\sim$3,500 km.  Such a large altitude range with our slant transit geometry indicates we are sensitive to pressures 
between 10's of mbar and $\sim$0.001 mbar.  
These limb pressures are in agreement with those found from \cite{Fortney03}, who analyzed the original \cite{Charby} 
medium resolution Na measurements.
At pressures higher than 30-50 mbar, our transmission spectra is optically thick to Rayleigh scattering by the 
abundant molecule H$_{2}$, providing a firm upper pressure limit.
%while at pressures
%below $\sim$0.001 mbar, Na is likely {\bf completely} ionized by stellar UV radiation \citep{Fortney03}.  
Such pressure restrictions, in addition to the strong narrow Na line cores, facilitate the necessity of a hot high-altitude 
temperature (above $\sim$1,500), which then naturally provides the observed altitude range due to an increased scale height.  
Our high altitude high limb temperature may point toward the atmospheric escape mechanism being felt, which necessitates large temperature
rises to perhaps 10,000 or 15,000 K in the exosphere \citep{Yelle,Vidal02,Vidal03}.

The highest altitudes probed here are dependent on the Na core, however, and thus have 
a correspondingly larger associated error.  The large error coupled with the possible Na depletion mechanisms
leaves open a wide range of possible upper atmospheric temperature profiles.
However, a high altitude temperature rise can be seen when analyzing the narrowest Na wavelength bands alone, sensitive
to high altitudes well above the condensation/ionization altitude, making the high-altitude temperature measurement
independent of the depletion mechanism.  The hot high-altitude temperature is illustrated in Fig. 4, which shows photometric Na measurements
from \cite{Charby} and \cite{Sing08a} overplotted with various constant abundance isothermal models.  
A single isothermal model fits through the 12 and 17 {\AA} bands, indicating that
the temperature and Na abundance do not change significantly between those altitudes ($\sim$2,000 to 2,200 km).
However, the absorptions measured in the 4.4 and 8.8 {\AA} bands, which probe higher in altitude ($\sim$2,700 to 3,700 km), 
rise much faster than the isothermal model fitting the lower altitudes of the 12 and 17 {\AA} bands, indicating that warmer temperatures
are needed.  

In the ionization Na depletion scenario, there are less temperature constraints for the middle atmosphere compared to the Na condensation model. 
However, limb temperatures everywhere above $\sim$1,500 K are unlikely, and in contradiction with Spitzer measurements of infrared brightness temperature.  
T-P profiles applicable to Na ionization models have less severe temperature rises compared to the condensation model, 
though a hot high-altitude temperature is still necessary and at least a modest temperature rise probable.
%However, the temperature change can be determined and is estimated to increase by 900$\pm$250 K over 1,400$\pm$200 km from the 12 and 4.4 {\AA} measurements alone.

A similar high altitude temperature rise will also be likely necessary
to explain recent measurements of another hot-Jupiter, HD189733b.  The Na core in that hot-Jupiter
reaches $\sim$4,950 km above the average optical transit altitude \citep{Redfield, Pont07}.
As the atmospheric scale height of HD189733b is smaller than HD209458b at the same temperature, 
a more extreme high altitude temperature rise is likely necessary, however, to fully explain the 
existing measurements.

\subsection{Comparison with Spitzer anti-transit data}

A stratospheric temperature inversion has been used to interpret Spitzer infrared secondary
eclipse measurements of HD209458b \citep{Knut08}, 
thought to be caused by an absorber at optical wavelengths 
on the day-side of the planet \citep{Burrows07, Fortney07}.  
This optical absorber has been postulated to be clouds or molecules TiO 
and VO \citep{Hubeny03, Burrows07, Fortney07}, with a low albedo 
ruling out some species, though not TiO or VO \citep{Rowe,Burrows07}. 

Comparing our limb T-P profile to the day-side profile from \cite{Burrows07}
(see Fig. 3), the atmospheric pressures probed are similar, between 0.01 and 30 mbar,
despite our slant transit viewing geometry \citep{Fortney05}.
The atmosphere in the infrared has a higher opacity,
due to strong molecular transitions from species such as water \citep{Burrows07},
placing the infrared surface of the planet viewed at normal angles at similar
altitudes as those probed in slant geometry in the optical at the limb. 

The high altitude temperature inversion we observe at the limb below pressures of $\sim$1 mbar 
is separate from that of the dayside and occurs at a much higher altitude. 
However, the altitude of TiO and VO we find from both models, largely confined to pressures above $\sim$10 mbar, 
closely matches the model presented by \cite{Burrows07}, 
who placed the stratospheric absorber at pressures of $\sim$25 mbar and imposed heat 
redistribution between pressures of 10 and 100 mbar.  The probable presence of TiO and VO in our spectrum along
with the altitude similarity implies TiO and VO 
are indeed likely to be the stratospheric day-side temperature inversion instigators \citep{Fortney07}.  
As our fit of TiO and VO is based largely on low resolution data, a more robust detection and interpretation of TiO/VO will 
likely require higher resolution {\it HST} spectra than those used here,
such that multiple specific molecular bandheads could be observed.

Some form of heat redistribution at or near our low-altitude pressure ($\sim$30 mbar) would likely be needed to 
explain the high temperature we find there, as the temperatures there are significantly above that predicted by
radiative equilibrium \citep{Showman08}.  
Initial Spitzer mid-infrared measurements 
of the brightness temperature as a function of orbital
phase have revealed that at least 30\% of the energy from the stellar flux must be 
redistributed to the night side to explain their null results, implying efficient heat redistribution \citep{Cowan}. 
Comparing with models by \cite{Burrows07,Burrows07b} with
higher P$_{n}$s ($>$0.35), night
side temperatures in the region of imposed heat distribution (corresponding to our pressure at $z_{s}$) 
can have larger temperatures than the day side, perhaps explaining our high temperature at $z_{s}$.  
Alternatively, systematic pressure scale differences between the
two studies may indicate that our transmission spectra instead probes the top of the stratospheric temperature inversion,
giving a potential explanation of the high temperature at $z_{s}$ as those
altitudes could already be warm.  
The minimum temperature we find, however, is in general agreement with radiative equilibrium, perhaps implying the
approximation is valid at those pressures around 1 mbar.
Such scenarios can be further distinguished in future studies,
with tighter constraints on the average infrared and optical opacities, additional infrared observations, along 
with improved global circulation models.  
The presence of a temperature inversion, TiO and VO along with heat redistribution 
would seem to currently place HD209458b in a 
intermediate hot-Jupiter class between the pM and pL classes as proposed by \cite{Fortney07}.

\section{Discussion}

The atmospheric temperatures we find, within the condensation scenario, and presence of TiO and VO can also explain other HD209458b
observational features.  For instance, the low middle atmospheric
temperature and the presence of TiO and VO would also explain the
lack of observed potassium (K), originally predicted to be a major contributor to the optical
opacity.   At the lower temperatures of the middle atmosphere, K condenses into potassium chloride
allowing KCl to become a dominant K-bearing gas.  Condensed K would then naturally deplete the atmosphere
of atomic K, as seen here with Na, leading to reduced signatures.
In the lower warmer atmosphere, where wide atomic
K line wings would be observable even at low resolution, TiO and VO likely make up the
surrounding continuum further masking the signature.  
If Na is instead ionized, the minimum limb temperature is higher and ionization of K, which is easier than Na to
ionize, likely depletes neutral K.  
If present in the atmosphere, a positive detection of atomic K 
in HD209458b will probably require high resolution ground-based or medium resolution space-based 
data to detect the K line core.  
Further detections or upper-limits of other atomic species which have different ionization potentials and condensation
temperatures can offer clues in future studies when distinguishing between the two depletion scenarios.

The Na mixing ratio we find for the lower atmosphere is about two times
solar while the abundance in the middle and upper atmosphere is found
to be 0.2 times solar or less, thus our models have decreased Na
abundance without the need for a global deficiency of Na.
In chemical equilibrium, Na$_2$S condensation effectively removes all the Na below the condensation
temperature, since the protosolar Na abundance is $\sim$13\% that of S, limiting 
the amount of Na$_2$S which can condense directly to the amount of Na.  
Our condensation model T-P profile crosses the condensation curve again at higher altitude, making
the altitudes of our minimum temperature a ``cold trap'' for Na.  It is currently unclear
how the 2$^{nd}$ crossing of the condensation curve could affect the Na abundance, as the
S/N limits us to fit only a single value for the whole upper Na abundance.

Assuming equilibrium chemistry, our temperature profile
would also result in less CO in the middle atmosphere, as our
temperature profile at that level is closer to the CO = CH$_4$ curve
which would result in an enhancement of CH$_4$ and H$_2$O over that of
CO and H$_2$ (see Fig. 2).  This is consistent with the lack of CO
detection in HD209458b during transit \citep{Deming}, although deviations from
equilibrium have been shown to push the CH$_4$/CO ratio to small
values for the pressures probed here of less than 1
bar \cite{Cooper06}.

Atmospheric data resulting from transit measurements represent the
average atmospheric properties across the observed limb of the planet,
which would include the morning, evening, and pole terminators.  As
previously recognized by \cite{Iro05}, temperature and structure differences could result
in differing transit signatures between the morning and evening terminators.
Longitudinal differences might also play a role, though the transit
geometry generally limits significant opacity contributions to regions of the terminator itself, as
the density drops rapidly from the terminator along the transit line of sight \citep{Hansen08}.
A T-P profile decreasing in temperature and pressure with increasing altitude would further 
mitigate longitudinal differences, corresponding to pressures above 1 mbar in our condensation model fit.
Efficient heat redistribution \citep{Cowan} implies all these differences 
could be minimal, though future Spitzer observations and the next
generation of global climate models would be able to better quantify the differences.
Given the size of our temperature uncertainties, limb differences of even several hundred
degrees would have a negligible impact on our resulting model fit.

Primary transit data alone is currently not able to give detailed knowledge on atmospheric
differences between the morning and evening terminators.  Such transit signatures
would appear during ingress and egress, where the transit light curve is less
sensitive to changes in planetary radii and more sensitive to errors in limb-darkening.
Nevertheless, as our Na~D line cores remain relatively narrow in the medium
resolution data, sensitive to the middle and upper atmospheres, a solar abundance
of Na on the potentially warmer evening terminator in the middle and upper atmospheres
is ruled out.  For the condensation model, this would indicate that Na settles from the middle
atmosphere and remains at lower altitudes, presumably at the Na$_2$S
condensation altitude.  Condensed Na$_2$S would have little impact on anti-transit
observations and the observational constraint of a low
albedo \citep{Rowe}.  On the day side, the hotter temperatures would
sublimate Na$_2$S back into their atomic constituents, and are
likely transparent in normal viewing geometry anyway \citep{Fortney05}.  

\section{Conclusions}

As seen here, fitting primary transit transmission spectra can give detailed atmospheric information
over large altitude ranges, and thus large pressure regimes, where different physical processes
may be important.  In our case, strong sodium absorption coupled with a relatively 
clear atmosphere provides this large altitude range and reveals the pressure where Na depletion via condensation or ionization takes place. 
The depletion level naturally provides
a contrast level between the warmer deeper atmosphere, likely sensitive to global circulation,
and a cooler middle atmosphere where many atomic and molecular species have condensed out and/or 
are depleted via ionization.  
As strong opacity in the core of the Na line permits measurements up to very low pressures, allowing Na
to be sensitive to altitudes perhaps as high as the lower thermosphere.  The slant transit geometry and high 
associated column density restricts the lowest altitudes probed, however, limiting this study to 
altitudes at and above the stratosphere.

Interpreting transit spectra becomes easier with increased wavelength coverage and increased
resolution.  For HD209458b, low resolution data spanning the optical reveals a planetary continuum 
comprised of H$_2$ Rayleigh scattering, sodium absorption, and TiO/VO giving a self consistent picture
of a hot atmosphere at pressures around $\sim$10 mbar.  This result is consistent with infrared
Spitzer emission features and a temperature inversion at those altitudes.  The medium resolution data places
the detection of Na absorption on firm ground, as there is sufficient S/N and resolution to both resolve the Na
doublet and observe the depletion level.  This overall atmospheric picture is not possible without data
from all three spectroscopic gratings and highlights the importance of future observations at other wavelengths
and resolutions.  

Transit transmission spectra have become a powerful method to probe the abundances of atomic and molecular species
for hot-Jupiters, as seen here for HD209458b and in \cite{Swain08} for HD189733b.  Both Jupiter and Saturn are enhanced in metals
relative to the sun, as would seem to be the case here for HD209458b having a sodium abundance about twice that of it's parent star.
Such studies can begin to answer the question of whether or not all hot-Jupiters are enriched in heavy elements,
shedding light on giant planet formation. 

After the next servicing mission, {\it HST} will be capable of extending the transmission 
spectra of HD209458b from the UV to NIR, potentially linking important physical atmospheric 
processes such as the altitude dependence of UV/NUV energy deposition as well as abundance and altitudes of other 
atomic and molecular species.
As currently a patchwork of such observations exist for the two ideal targets HD209458b and HD189733b, future observations
should focus on covering wavelength and resolution gaps, which can lead to both theoretical and observational 
links between the separate studies.

\acknowledgments
D.K.S. is supported by CNES.  This work is based on observations
with the NASA/ESA Hubble Space Telescope, obtained at the
Space Telescope Science Institute (STScI) operated by AURA, Inc.  
Support for this work was provided by NASA through a grant from the STScI.
We thank G. Tinetti, F. Pont, G. H\'{e}brard, R. Ferlet, F. Bouchy and N. Allard for discussions and insight.
We also like to especially thank our referee Dr. Jonathan Fortney for his insightful comments and valuable suggestions,
especially concerning sodium ionization.

\clearpage
%========================
%==========
%Figure 11.
\begin{figure*}
  \vspace{1cm}
  \plotone{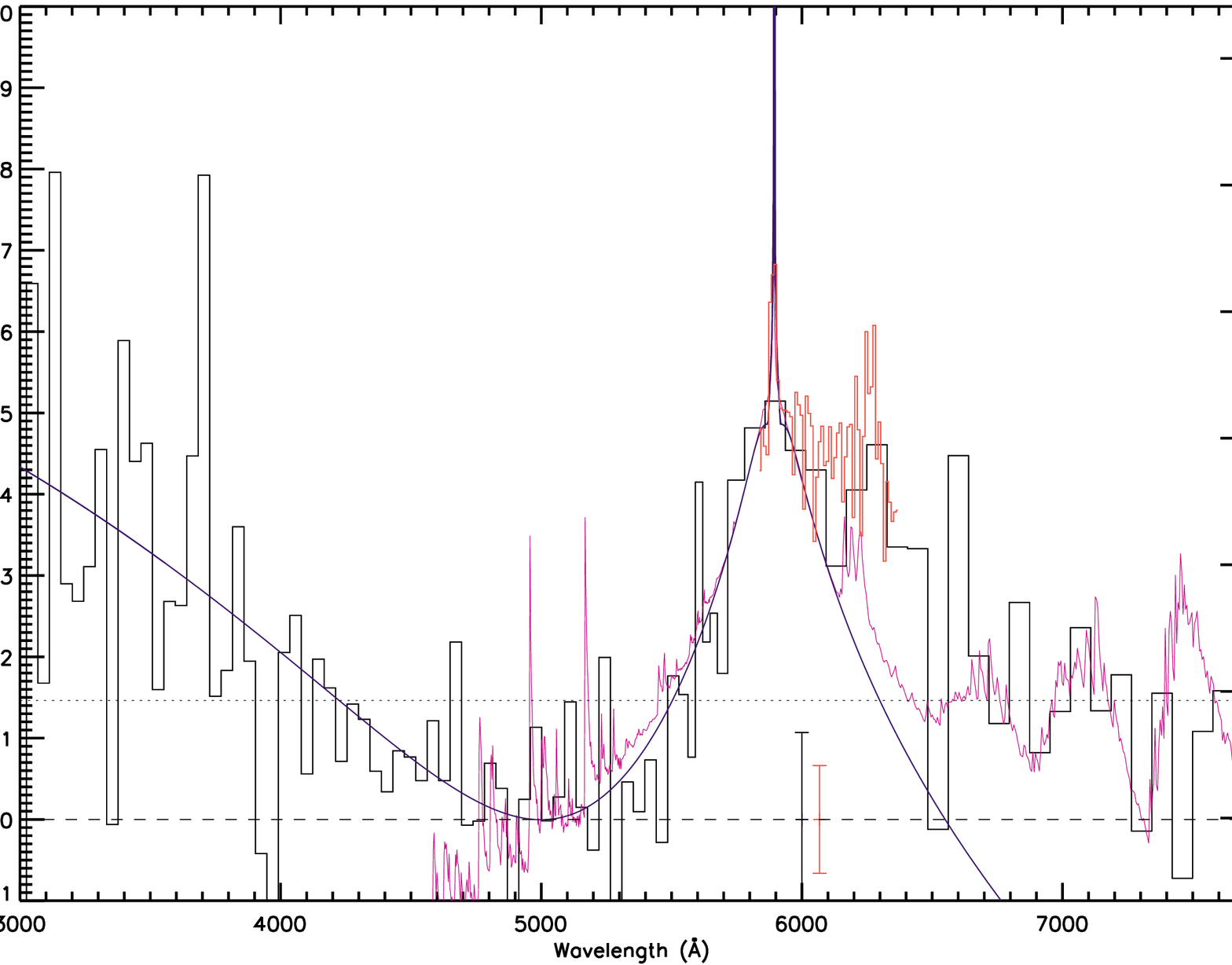}
  \figcaption{Transit Na~D absorption from \cite{Sing08a}. Plotted 
is the planetary absorption percentage compared to the minimum
absorption level referenced at 5,000{\AA} (1.440\%).  In black is the low resolution
data binned over 16 pixels and overplotted in red is the medium
resolution data binned over 18 pixels (10 \AA).  Representative error bars for the
binned data are also plotted (also see Fig. 10 \& 11 of \citealt{Sing08a}) as well as  
the mean planetary radius from \cite{Knut} (dotted line).
Plotted in blue is our transit absorption model
which contains opacity from atomic Na, Rayleigh scattering, and
contains upper-atmospheric Na depletion.  The model in purple includes Na, TiO, and VO illustrating
the wavelengths where those features are prominent.  
Higher absorption values
indicate higher atmospheric altitudes, with the right y-axis
labeled as the altitude above our minimum absorption depth, assuming
the system parameters of \cite{Knut}.  
The Na~D line cores can easily
be seen in the zoomed plot of Fig. 2.  The narrow Na~D1 and D2 cores
extend above a wide plateau of Na absorption indicating a low amount of
Na is present in the middle atmosphere compared to the lower.  The
drastic abundance change, along with the pressure and temperature
where the change occurs, indicates that Na is condensing into sodium
sulfide on the night side of the planet. 
}
\end{figure*}
%========================
%Figure 11.
\begin{figure}
 \vspace{1cm}
  \plotone{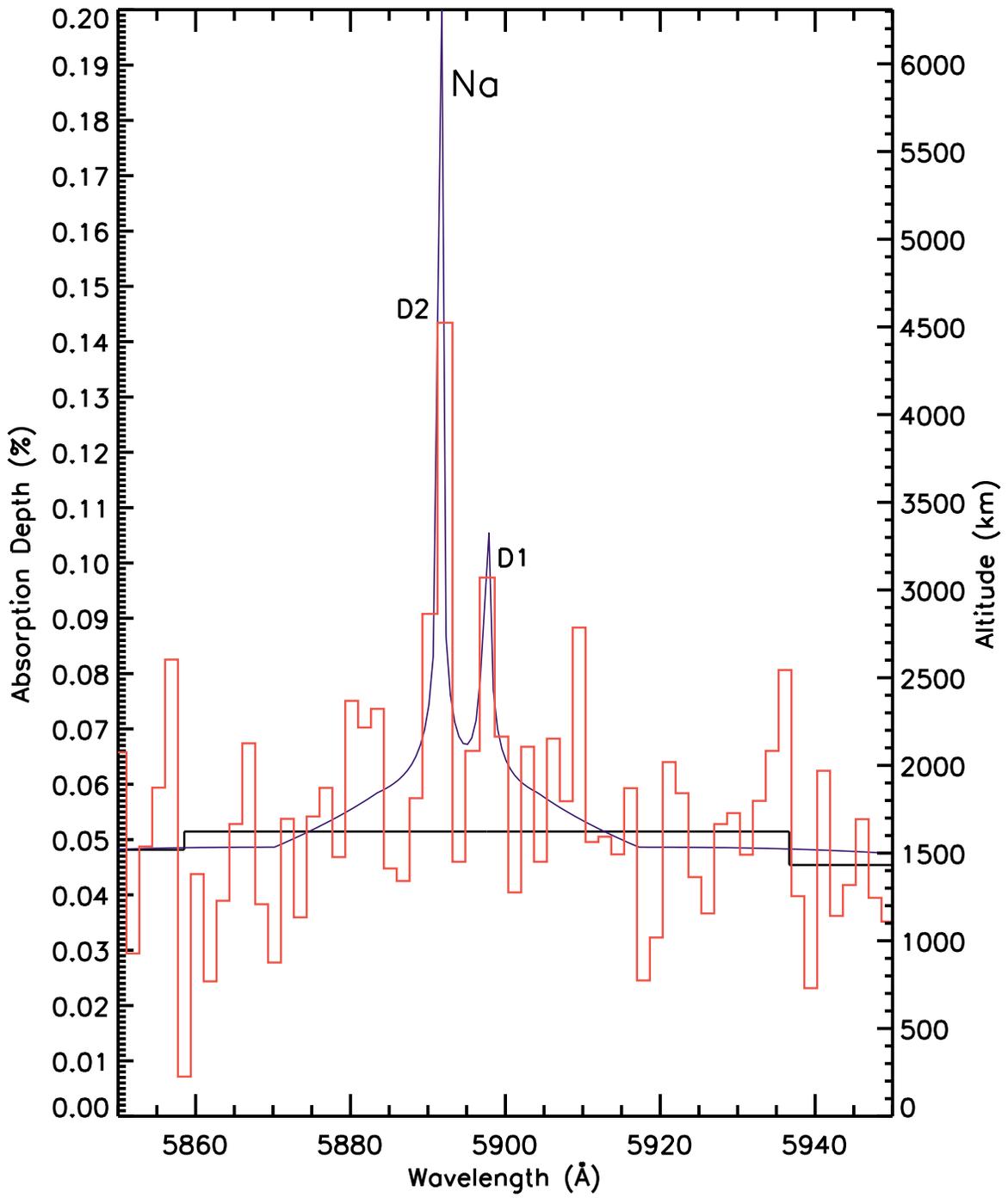}
  \figcaption{Same as in Fig. 1, but zoomed on the Na region with the medium resolution data
binned over 3 pixels (1.7 \AA).  The Na~D doublet is easily resolved and reaches up to an
average altitude of $\sim$3,500 km.}

\end{figure}
%========================
%========================
\begin{figure}
 \vspace{0.2cm}
  \plotone{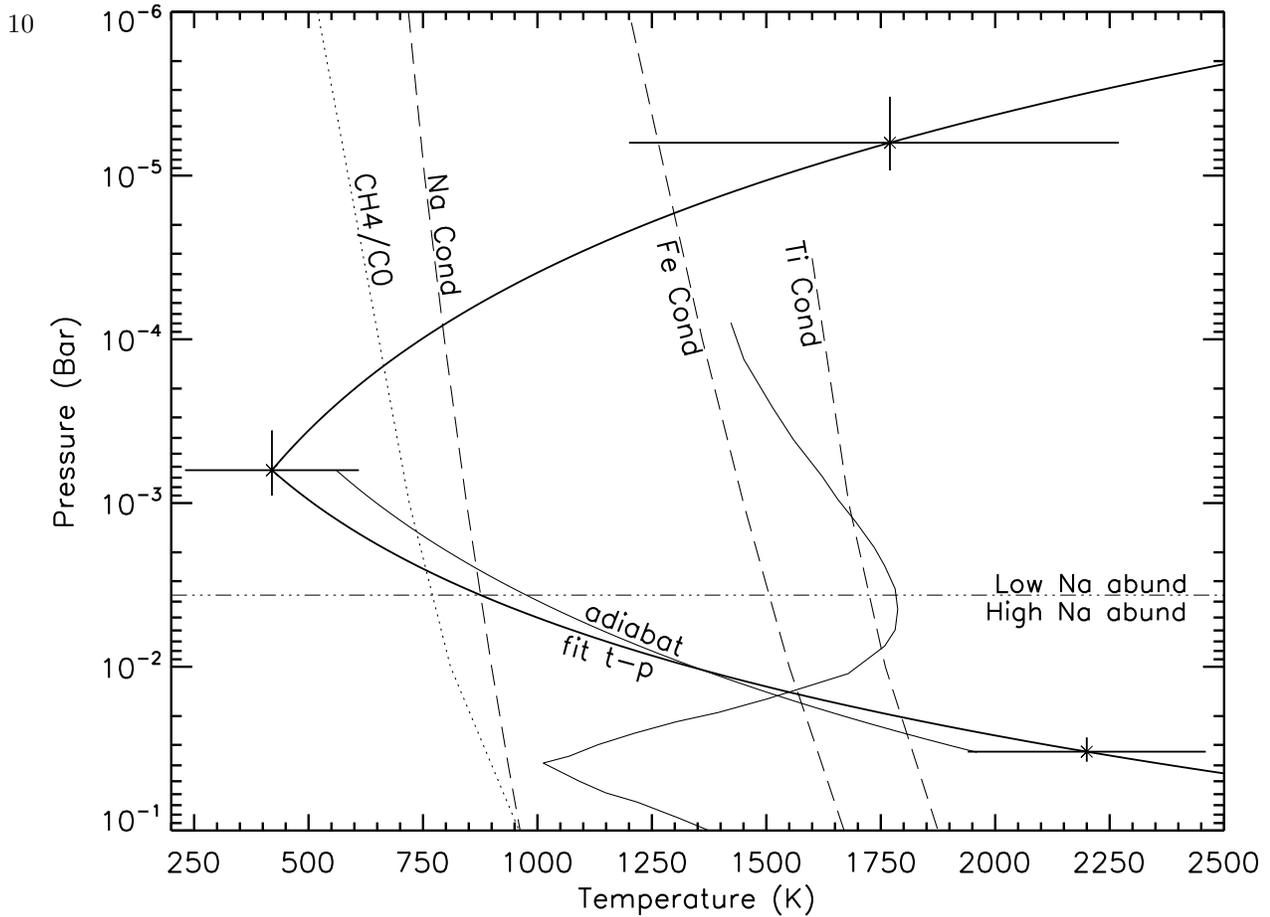}
  \figcaption{Limb temperature-pressure atmospheric profile of HD209458b for the sodium condensation model.
Plotted (thick line, three points with error bars) is our best fit terminator T-P profile, 
while (thin) is a day-side inverted profile
from \cite{Burrows07}.  Also shown are the
chemical equilibrium condensation curves of iron, TiO, and sodium (dashed
lines). The horizontal line represents the pressure as which the T-P
profile crosses the sodium condensation curve.  Above the condensation
curve, at high altitudes and low pressures, Na is depleted compared to
lower altitudes. The CH$_4$/CO line represents the profile where equal
amounts of CO and CH$_4$ are expected assuming chemical equilibrium.
The whole data set with low and medium
resolution observations are fitted with a total of 7 free
parameters, resulting in the plotted T-P profile and sodium abundance
below and above the sodium condensation temperature.  
A fit imposing an adiabatic lower T-P profile is also shown.
The sodium
abundance above the Na$_2$S condensation level is found to be
significantly lower than the sodium abundance below the condensation
level (the ratio of both abundances is significantly above 1 at
3-$\sigma$ level); this reveals the presence of Na$_2$S condensation.}

\end{figure}

%========================
\begin{figure}
 \vspace{0.2cm}
  \plotone{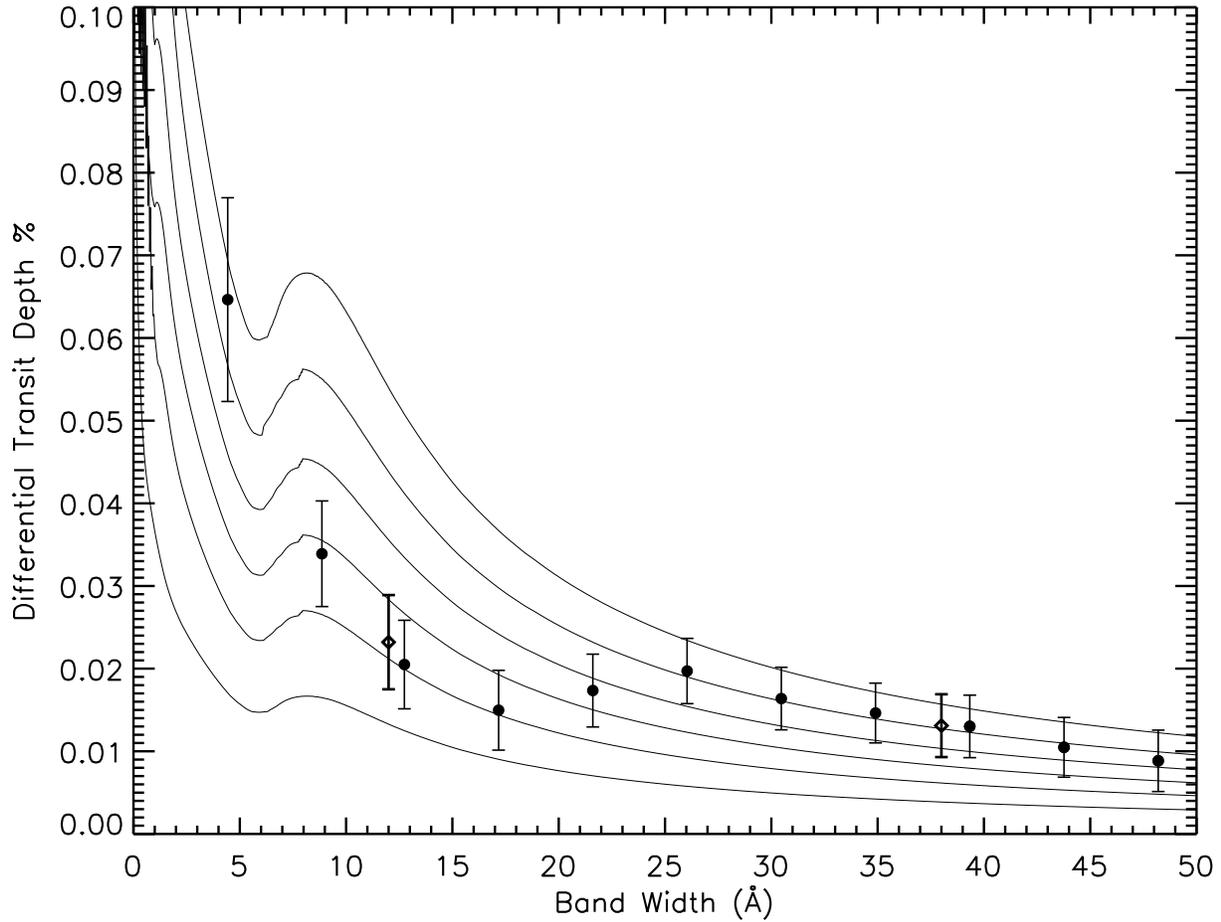}
  \figcaption{Differential spectrophotometric measurements of atmospheric sodium 
absorption from (diamonds) Charbonneau et al. (2002) and (circles) Sing et al. (2008) using the HST/STIS medium
resolution data.  Also plotted are the Na absorption predictions from isothermal models with temperatures (bottom to top)
of 500, 750, 1000, 1250, 1500, and 2000 K and constant sodium abundance.  The isothermal models illustrate the necessity
for a high-altitude temperature increase in the narrowest two photometric bands (4.4 and 8.9 {\AA}) which probe
the highest altitudes.}

\end{figure}
%========================

\end{document}